\documentclass{article}

\usepackage{spconf}
\newif\ifFullVersion
\usepackage{cite}
\usepackage{amsmath,amssymb,amsfonts}
\usepackage{graphicx}
\usepackage{textcomp}
\usepackage{xcolor}
\usepackage{tikz}
\usepackage{xcolor}
\usetikzlibrary{positioning, arrows.meta, shapes.geometric, fit, backgrounds, shadows}
\usepackage{multirow}
\usepackage[table,xcdraw]{xcolor}
\usepackage[ruled,linesnumbered,vlined]{algorithm2e}

\usepackage{multirow} 
\usepackage{makecell} 

\definecolor{lightblue}{RGB}{173, 216, 230}
\definecolor{lightorange}{RGB}{255, 218, 185}
\definecolor{lightgreen}{RGB}{144, 238, 144}
\definecolor{lightyellow}{RGB}{255, 255, 224}
\definecolor{lightpink}{RGB}{255, 182, 193}
\definecolor{lightpurple}{RGB}{221, 160, 221}
\definecolor{mediumblue}{RGB}{135, 206, 235}
\definecolor{mediumgreen}{RGB}{152, 251, 152}
\definecolor{darkblue}{RGB}{70, 130, 180}
\definecolor{darkpurple}{RGB}{138, 43, 226}

\tikzset{
    block/.style={
        rectangle, 
        draw, 
        thick, 
        rounded corners=4pt,
        minimum height=2.5cm, 
        minimum width=1.2cm,
        text width=1cm, 
        align=center, 
        font=\tiny\bfseries,
        inner sep=0.5pt,
        text depth=0pt,
        text height=0.2cm
    },
    wideblock/.style={
        rectangle, 
        draw, 
        thick, 
        rounded corners=4pt,
        minimum height=2.5cm, 
        text width=1.8cm,
        align=center, 
        font=\tiny\bfseries,
        inner sep=0.2pt
    },
    smallblock/.style={
        rectangle, 
        draw, 
        thick, 
        rounded corners=3pt,
        minimum height=0.8cm, 
        text width=1.3cm, 
        align=center, 
        font=\scriptsize,
        inner sep=1pt
    },
    circleblock/.style={
        circle,
        draw,
        thick,
        minimum size=0.6cm,
        align=center,
        font=\scriptsize,
        inner sep=1pt
    },
    arrow/.style={-{Stealth[length=2mm, width=1.5mm]}, thick},
    dashedarrow/.style={-{Stealth[length=2mm, width=1.5mm]}, thick, dashed, red},
    estimatorbg/.style={
        fill=lightblue!20, 
        draw=darkblue, 
        rounded corners=6pt,
        inner sep=8pt
    },
    monitorbg/.style={
        fill=lightpurple!15, 
        draw=darkpurple, 
        rounded corners=6pt,
        inner sep=8pt
    }
}

\def\BibTeX{{\rm B\kern-.05em{\sc i\kern-.025em b}\kern-.08em
    T\kern-.1667em\lower.7ex\hbox{E}\kern-.125emX}}

\usepackage{acronym}
\usepackage[belowskip=-10pt,aboveskip=1.8pt]{caption}
\usepackage{standalone}
\usepackage{subcaption}
\usepackage{tikz}
\usepackage{url,enumitem, cite}
\usepackage{verbatim}
\usepackage[bookmarks,colorlinks]{hyperref}
\usepackage{soul, xcolor}
\usepackage{mathrsfs}
\usepackage[ruled,linesnumbered,vlined]{algorithm2e}
\usepackage{dsfont}
\SetKwInput{KwData}{\textbf{Init}}

\usepackage[all=normal,paragraphs=tight,floats=normal,mathspacing=normal,wordspacing=tight,charwidths=tight,mathdisplays=normal,leading=normal]{savetrees}
\usepackage{pgfplots}
\usepackage{pgfplotstable}
\usetikzlibrary{spy}
\pgfplotsset{compat=newest}
\usepackage{booktabs}

\usepackage{amsmath,amssymb,mathtools}
\usepackage{empheq} 
\newcommand{\myVec}[1]{{\boldsymbol{#1}}}
\newcommand{\myMat}[1]{{\boldsymbol{#1}}}
\newcommand{\mySet}[1]{\mathcal{#1}}

\let\oldnl\nl
\newcommand{\nonl}{\renewcommand{\nl}{\let\nl\oldnl}}

\newcommand{\update}[1]{\textcolor{black}{#1}}

\newcommand{\freqAmbi}[1]{\myVec{a}_{#1}(k)}
\newcommand{\hatfreqAmbi}[1]{\hat{\myVec{a}}_{#1}(k)}

\newcommand{\compressSTFT}[1]{\mathcal{H}(#1)}
\newcommand{\uncompressSTFT}[1]{\mathcal{H}^{-1}(#1)}

\acrodef{ai}[AI]{artificial intelligence}
\acrodef{adaln}[AdaLN]{adaptive linear norm}
\acrodef{dl}[DL]{deep learning}
\acrodef{bs}[BS]{base station}
\acrodef{dnn}[DNN]{deep neural network}
\acrodef{cnn}[CNN]{convolutional neural network}
\acrodef{mlp}[MLP]{multi-layer perceptron}
\acrodef{snr}[SNR]{signal-to-noise ratio}
\acrodef{awgn}[AWGN]{additive white Gaussian noise} 
\acrodef{ml}[ML]{machine learning} 
\acrodef{sgd}[SGD]{stochastic gradient descent} 
\acrodef{gd}[GD]{gradient descent} 
\acrodef{mse}[MSE]{mean-squared error}
\acrodef{msie}[MSIE]{mean-squared innovation error}
\acrodef{rmse}[RMSE]{root mean squared error}
\acrodef{mspe}[MSPE]{mean-squared periodic error}
\acrodef{rmspe}[RMSPE]{root \ac{mspe}}
\acrodef{mle}[MLE]{maximum likelihood estimation}
\acrodef{snr}[SNR]{signal-to-noise ratio}
\acrodef{admm}[ADMM]{alternating direction method of multipliers}
\acrodef{aoa}[AoA]{Angle of Arrival}
\acrodefplural{aoa}[AoAs]{Angles of Arrival}
\acrodef{em}[EM]{electromagnetic}
\acrodef{cmos}[CMOS]{complementary metal-oxide semiconductor}
\acrodef{ula}[ULA]{uniform linear array}
\acrodef{em}[EM]{Electromagnetic}
\acrodef{doa}[DoA]{direction of arrival}
\acrodef{music}[MUSIC]{MUltiple SIgnal Classification}
\acrodef{esprit}[ESPRIT]{Estimation of signal parameters via rotational invariance techniques}
\acrodef{evd}[EVD]{eigenvalues decomposition}
\acrodef{sps}[SPS]{spatial smoothing}
\acrodef{iid} [i.i.d] {Independent and identically distributed}
\acrodef{ls}[LS]{Least Square}
\acrodef{relu}[ReLu]{Rectified Linear Unit}
\acrodef{crb} [CRB] {Cramér–Rao Bound}
\acrodef{ccrb} [CCRB] {Conditional Cramér–Rao Bound} 
\acrodef{vqvae}[VQ-VAE]{vector quantized variational autoencoder}
\acrodef{rm}[Root-MUSIC]{Root-MUSIC}
\acrodef{esprit}[ESPRIT]{Estimation of Signal Parameters via Rotational Invariance Techniques}
\acrodef{drm}[DR-MUSIC]{Deep Root-MUSIC}
\acrodef{ssn}[SubspaceNet]{Subspace Net}
\acrodef{rssn}[Remote SubspaceNet]{Remote Subspace Net}
\acrodef{mbdl}[Model Based Deep Learning]{Model Based Deep Learning}
 
\acrodef{kf}[KF]{Kalman filter} 
\acrodef{doa}[DOA]{direction-of-arrival} 
\acrodef{sh}[SH]{spherical harmonics} 
\acrodef{sgm}[SGM]{score-based generative model} 
\acrodef{sde}[SDE]{stochastic differential equation} 
\acrodef{au}[AU]{Ambisonics upscaling} 
\acrodef{foa}[FOA]{first-order Ambisonics} 
\acrodef{hoa}[HOA]{high-order Ambisonics} 
\acrodef{stft-sdr}[STFT-SDR]{STFT signal-to-distortion ratio} 
\acrodef{stft-si-sdr}[STFT-SI-SDR]{STFT scale invariant signal-to-distortion ratio}

\acrodef{ve}[VE]{Variance Exploding} 
\acrodef{ode}[ODE]{ordinary differential equation}
\acrodef{dps}[DPS]{Diffusion Posterior Sampling} 
\acrodef{atf}[ATF]{array transfer function} 
\acrodef{sht}[SHT]{Spherical Harmonics Transform} 
\acrodef{arir}[ARIR]{Ambisonics room impulse response} 
\acrodef{adeps}[ADEPS]{Ambisonics Diffusion Encoding via Posterior Sampling}

\acrodef{tf}[TF]{time-frequency}
\acrodef{stft}[STFT]{short-time Fourier transform}
\acrodef{istft}[ISTFT]{inverse STFT}
\acrodef{ncsn}[NSCN++]{noise-conditioned score-matching network}
\acrodef{rd}[RD]{reverse diffusion}
\acrodef{ald}[ALD]{annealed Langevin dynamics}
\acrodef{mushra}[MUSHRA]{MUltiple Stimuli with Hidden Reference and Anchor}
\acrodef{pwd}[PWD]{plane wave decomposition}
\acrodef{cs}[CS]{compressed sensing}
\acrodef{ild}[ILD]{interaural level difference}
\acrodef{ic}[IC]{interaural coherence}

\acrodef{cs}[CS]{compressed sensing}
\acrodef{si-sdr}[SISDR]{signal invariant signal distortion ratio}
\acrodef{erb}[ERB]{equivalent rectangular bandwidth}
\acrodef{film}[FiLM]{Feature-wise Linear Modulation}

\setstcolor{blue}

\begin{document}

\title{Array-Agnostic
Ambisonics Encoding via Diffusion Posterior Sampling
}

\name{Amit Milstein, Nir Shlezinger, and Boaz Rafaely
\address{\small ECE School Ben-Gurion University of the Negev, Be’er-Sheva, Israel (e-mail: amitmils@post.bgu.ac.il; \{nirshl; br\}@bgu.ac.il).
}
\vspace{-5mm}
}

\maketitle

\begin{abstract}
Spatial audio enhances user immersion by reproducing 3D sound fields, with Ambisonics being a widely adopted representation. While Ambisonics is theoretically independent of the recording setup, practical microphone arrays introduce  hardware-dependent encoding artifacts. Moreover, existing data-driven solutions lack flexibility, as they are typically restricted to fixed array geometries. To overcome these limitations, we propose \ac{adeps}, a generative framework that explicitly embeds the physical acquisition model into the inference process. By leveraging this formulation, \ac{adeps} effectively compensates for array-specific distortions while enabling zero-shot encoding across arbitrary array topologies. We train the underlying generative prior in an unsupervised manner solely on target Ambisonic representations. Extensive evaluations across diverse simulated and real microphone arrays demonstrate that \ac{adeps} consistently outperforms both traditional linear and parametric baselines in spatial fidelity and spectral quality.
\end{abstract}


\acresetall

\vspace{-0.2cm}
\section{Introduction}
\label{sec:intro}
Spatial audio enables realistic delivery of sound by synthesizing the three-dimensional properties of sound sources and acoustic environments. It thus forms the foundation for immersive audio experiences 
\cite{hacihabiboglu2017perceptual}.
Among various spatial audio representations, Ambisonics has emerged as a dominant format due to its theoretical independence from recording configurations \cite{zotter2019ambisonics}. In practical applications, however, Ambisonics representations must be extracted from physical microphone arrays. This encoding step is  limited by the array's physical properties, where sound-field sampling, microphone positioning, and acoustic diffraction introduce severe array-dependent encoding constraints \cite{rafaely2005analysis}. Suppressing physical encoding artifacts, particularity in an array-agnostic manner, remains a core challenge in  spatial audio acquisition.


Classical model-based approaches for Ambisonics encoding include signal-independent methods that formulate the encoding task as an inverse problem using a theoretical or measurement-based steering matrix~\cite{politis2017comparing,moreau20063d,poletti2005three,bastine2022ambisonics,gayer2025ambisonics}. These methods often exhibit severe artifacts due to matrix ill-conditioning and underdetermined systems. Alternatively, signal-dependent parametric methods~\cite{mccormack2022parametric,politis2018compass} attempt to dynamically resynthesize the sound field. While these models provide clear physical interpretability, they tend to underperform in reverberant environments due to errors in \ac{doa} estimation, difficulty in separating closely spaced sources, and the challenge of accurately modeling complex reflection patterns beyond sparse directional and residual components.

Recent data-driven methods utilize neural networks to overcome the limitations of model-based encoding approaches. Various neural architectures were shown to learn accurate encoding mappings specific array geometries~\cite{heikkinen2024neural,deppisch2026residual,qiao2025neural}. To support  different configurations, Heikkinen et al. ~\cite{heikkinen2025gen} proposed conditioning the network on microphone positions; yet, this approach remains restricted to a fixed number of channels and neglects array-specific frequency dependent characteristics. The follow-up work ~\cite{heikkinen2026beyond} introduced an encoder–decoder architecture using cross-attention between captured signals and \acp{atf} to take in account the array's physical properties. However, because the architecture lacks channel-flexibility, adapting it to an array with a different microphone count requires complete retraining.

To overcome these limitations, we introduce a novel encoding framework for arbitrary microphone arrays termed {\em \ac{adeps}}. To achieve array-agnostic encoding, we embed the physical signal acquisition model into a \ac{dps} generative pipeline~\cite{chung2022diffusion}. The generative prior is trained  on ideal, uncorrupted Ambisonics representations, such that the learned model remains  agnostic to specific array geometries and their associated artifacts. At inference time, \ac{adeps} leverages this learned prior alongside explicit knowledge of the physical array degradation model (which is dependent on the  \ac{atf}), enforcing measurement consistency against a linearly encoded signal throughout the generative refinement process. We show that on reverberant speech signals, our method consistently outperforms existing array-agnostic and array specific approaches in signal fidelity across a variety of microphone arrays.

\vspace{-0.4cm}
\section{Signal Model and Preliminaries}
\label{sec:System Model and Problem Formulation}  
\vspace{-0.4cm}
  
\subsection{Signal Model}
\label{subsec:Signal Model}
Ambisonics is a common spatial audio format rooted in the spatial Fourier transform on the unit sphere \cite{zotter2019ambisonics}. 
This format parametrizes a sound field  by a set of frequency-dependent Ambisonics coefficients, $a_{nm}(k)$ (where $k$ is the wave number), which act as the weights for an infinite series of orthogonal basis functions known as \ac{sh} of order $n$ and degree $m$.
    
When this sound field is sampled by an arbitrary array comprising $Q$ omnidirectional microphones, with 
the position of the $q$-th microphone  specified by the spherical coordinates 
$\myVec{r}_q = (r_q, \theta_q, \phi_q)$, the acoustic pressure captured by the $q$-th microphone 
can be modeled as a function of these coefficients \cite{rafaely2015fundamentals}
\begin{equation}\label{eq:microphone_acquisition_model}
    p_q(k) \!=\! \sum_{n=0}^{\infty} \sum_{m=-n}^{n} \!b_n(k, r_q) Y_n^m(\theta_q, \phi_q) a_{nm}(k) \!+\! \nu_q(k).
\end{equation}
In \eqref{eq:microphone_acquisition_model}, $b_n(k, r_q)$ is the order-dependent radial function; $Y_n^m(\theta_q, \phi_q)$ denotes the real-valued \ac{sh} function; and  $\nu_q(k)$ is additive measurement noise,  assumed to be uncorrelated with 
the Ambisonics coefficients and i.i.d. across all microphones.

Due to the rapid decay of the radial functions $b_n(k, r_q)$ for orders exceeding the spatial Nyquist limit ($n > \lceil k r_q \rceil$), the infinite series in \eqref{eq:microphone_acquisition_model} can be practically truncated to an effective Ambisonics order, $N_{\text{eff}}$ \cite{rafaely2015fundamentals}. This truncation yields a finite number of Ambisoncs coefficients, allowing the sound field acquisition across the array to be compactly modeled in matrix notation  as 
\begin{equation}\label{eq:matrix_signal_acquisition}
    \myVec{p}(k) = \myVec{V}(k) \freqAmbi{N_{\text{eff}}} + \myVec{\nu}(k) \in \mathbb{C}^Q,
\end{equation}
where $ \freqAmbi{N_e}= [a_{00}(k),...,a_{N_{\text{eff}}N_{\text{eff}}}(k)]^T$ stacks the $(N_{\text{eff}}+1)^2$ Ambisonics coefficients, $\myVec{\nu}(k)$ is the measurement noise vector, and the matrix $\myVec{V}(k) \in \mathbb{C}^{Q \times (N_e+1)^2}$ represents the modal steering matrix. The latter  can be further decomposed into
\begin{equation}\label{eq:decomposed_steering_matrix}
    \myVec{V}(k) = \myVec{Y}_{N_{\text{eff}}}\myVec{B}(k),
\end{equation}
in which $\myVec{B}(k)\in\mathbb{C}^{(N_{\text{eff}}+1)^2\times(N_{\text{eff}}+1)^2}$ is a diagonal matrix of radial functions and $\myVec{Y}_{N_{\text{eff}}}\in\mathbb{R}^{Q\times(N_{\text{eff}}+1)^2}$ is the \ac{sh} matrix.

\subsection{Problem Formulation}
\label{sec:ProblemFormulation}
 Ambisonics encoding~\cite{poletti2005three} refers to reconstructing Ambisonics coefficients from captured signals by a microphone array \eqref{eq:matrix_signal_acquisition}. Specifically, an encoder $\mathrm{E}$ transforms discrete microphone signals into their corresponding order $N_{\text{enc}}$ Ambisonics coefficients, i.e., 
\begin{equation}
    \hatfreqAmbi{N_{\text{enc}}} = \mathrm{E}\left( \myVec{p}(k), \myVec{V}(k) \right).
    \label{eq:encoding}
\end{equation}

 By~\eqref{eq:decomposed_steering_matrix}, the mapping \eqref{eq:encoding} can be formulated  as an inverse problem~\cite{politis2017comparing, heikkinen2024neural}. However, this  gives rise to several challenges:
\begin{enumerate}[label=$C_\arabic*$, leftmargin=*,series=challenges]
    \item \label{item:c1} \textbf{Noise Amplification:} At low frequencies ($k \to 0$), the entries of $\myVec{B}(k)$ for high orders $n$ become extremely small, making direct inversion ill-posed and  amplifies sensor noise.
    \item \label{item:c2} \textbf{Non-Uniform  Sensitivity:} Irregular or non-ideal microphone positions degrade the condition number of   $\myVec{Y}_{N_e}$, yielding 
    direction- and channel-dependent reconstruction errors.
    \item \label{item:c3} \textbf{High-Frequency Spatial Aliasing:} When the number of sensors holds $Q < (N_{\text{eff}}+1)^2$,  spatial sampling violates the  Nyquist criterion, introducing  aliasing  at higher frequencies. (ii) 
\end{enumerate}
While \ref{item:c1}-\ref{item:c3} are inherent to the Ambisonics encoding task, here we seek to overcome an additional challenge
\begin{enumerate}[resume*=challenges]
    \item \label{item:c4} \textbf{Array-Invariance:} 
    The encoder should be applicable across array   geometries and microphone counts $Q$.
\end{enumerate} 
Our objective is this to establish a spatial mapping operator that is applicable across different arrays. 
To cope with \ref{item:c1}-\ref{item:c4}, we assume access  to ideal, array-agnostic $N_p$-order Ambisonics 
coefficients during training, which we formalize as 
    $\mySet{D} = \big\{\myVec{a}_{N_p}^{(i)} \big\}_{i=1}^{|\mySet{D}|}$.
\vspace{-0.2cm}
\subsection{Preliminaries}\label{ssec:Preliminaries}
{\bf Linear Ambisonics Encoding} is a common encoding rule. It seeks a matrix $\myMat{E}(k)$ from which order-$N_{\text{enc}}$ Ambisonics are encoded as
$\hatfreqAmbi{N_{\text{enc}}} = \myVec{E}(k) \myVec{p}(k)$. Based on \eqref{eq:matrix_signal_acquisition}, it sets $\tilde{\myVec{E}}(k) \in \mathbb{C}^{(N_e+1)^2 \times Q}$ as the Tikhonov-regularized  pseudo-inverse~\cite{moreau20063d,poletti2005three}
\begin{equation}\label{eq:linear_encoder}
    \tilde{\myVec{E}}(k) = \myVec{V}^H(k) \left[ \myVec{V}(k) \myVec{V}^H(k) + \gamma^2 \myVec{I}_Q \right]^{-1},
\end{equation}
where $(\cdot)^H$ denotes the conjugate transpose, $\myVec{I}_Q$ is the $Q \times Q$ identity matrix, and $\gamma^2$ is a regularization hyperparameter.

Due to the physical constraint of the array having $Q$ microphones, the target encoding order $N_{\text{enc}}$ must satisfy $(N_{\text{enc}}+1)^2 \le Q$ to remain spatially resolvable \cite{rafaely2015fundamentals}. Consequently, the  encoding matrix $\myVec{E}(k) \in \mathbb{C}^{(N_{\text{enc}}+1)^2 \times Q}$ is formed by isolating the first $(N_{\text{enc}}+1)^2$ rows of $\tilde{\myVec{E}}(k)$. Although applicable to any  configuration (satisfying \ref{item:c4}), linear encoding is sensitive to \ref{item:c1}-\ref{item:c3}.

 \smallskip
 \noindent
{\bf Score-Based Generative Inverse Problems} are a family of emerging data-driven techniques for tackling inverse problems using  continuous-time diffusion models.
To formulate such methods, consider a forward diffusion process that  perturbs a clean data sample $\myVec{x}_0 \sim p_{\text{data}}$ into Gaussian noise over a  time horizon $\tau \in [0,\tau_{\max}]$. To reverse this process, one can solve the probability flow \ac{ode}~\cite{song2020score}
\begin{equation}\label{eq:ODE}
    d\myVec{x}_\tau = \left[ f(\myVec{x}_\tau,\tau) - \frac{1}{2} g(\tau)^2 \nabla_{\mathbf{x}_\tau} \log p_\tau(\myVec{x}_\tau) \right] \mathrm{d}\tau,
\end{equation}
where $\nabla_{\myVec{x}_\tau} \log p_\tau(\myVec{x}_\tau)$ denotes the score function of the marginal probability density at time $\tau$, while $f(\cdot)$ and $g(\cdot)$ represent the drift and diffusion coefficients, respectively.
Following~\cite{karras2022elucidating}, the score can be computed using a neural network $D_\myVec{\theta}(\myVec{x}_\tau, \sigma(\tau))$ that is trained to act as a denoiser at noise level $\sigma(\tau)$. By applying Tweedie's formula~\cite{robbins1992empirical}, the score function can be approximated as
\begin{equation}\label{eq:score_approx}
\nabla_{\myVec{x}_\tau} \log p_\tau(\myVec{x}_\tau) \approx {\big(D_\myVec{\theta}(\myVec{x}_\tau, \sigma(\tau)) - \myVec{x}_\tau\big)}/{\sigma(\tau)^2},
\end{equation}

Rather than generating samples, this approach can be used to solve inverse problems by leveraging learned complex priors. To this end,  assume that $\myVec{x}_0$ is observed through a measurement system modeled as $\myVec{y}=\mathcal{A}(\myVec{x}_0)+\myVec{\epsilon}$, where $\mathcal{A}$ is a known forward degradation function and $\myVec{\epsilon} \sim \mathcal{N}(\myVec{0}, \sigma^2_y \myVec{I})$. The \ac{ode} formulation can be used to sample from the posterior distribution $p(\myVec{x} | \myVec{y})$, which requires evaluating the conditional score function $\nabla_{\myVec{x}_\tau} \log p_\tau(\myVec{x}_\tau | \myVec{y})$ in \eqref{eq:ODE}.
\ac{dps}~\cite{chung2022diffusion} tackles this  using Bayes' rule to decompose the posterior score function into two distinct terms: 
\begin{equation}\label{eq:bayes_decomposition}
\nabla_{\myVec{x}_\tau}\! \log p_{\tau}(\myVec{x}_\tau | \myVec{y}) \!= \!\nabla_{\myVec{x}_\tau} \!\log p_{\tau}(\myVec{x}_\tau) \!+\! \nabla_{\myVec{x}_\tau} \!\log p_{\tau}(\myVec{y} |\myVec{x}_\tau ),
\end{equation}
where the first term is the unconditional prior score and the second is the conditional likelihood score. As the likelihood score is intractable for  $\tau > 0$, it is often approximated via the measurement error relative to the current MMSE estimate  \cite{chung2022diffusion, moliner2023solving}:
\begin{equation}\label{eq:likelihood_guidance}
\!\!\nabla_{\myVec{x}_\tau} \!\log p_{\tau}(\myVec{y} |\myVec{x}_\tau ) \!\approx\! -\eta(\tau) \nabla_{\myVec{x}_\tau} \!\|\myVec{y} \!-\! \mathcal{A}(D_\myVec{\theta}(\myVec{x}_\tau, \sigma(\tau)))\|_2^2,
\end{equation}
where $\eta(\tau)$ is a time-dependent step size.

\section{ADEPS}
\label{sec:method}
We next introduce \ac{adeps} which performs Ambisonics encoding on reverberant speech by using \ac{dps} to sample from the ideal Ambisonics distribution conditioned on the linearly encoded coefficients. 
All signals are  formulated in the \ac{stft} domain, and their  time-frame and frequency-bin indices are omitted  for brevity.

\subsection{DPS Formulation for Ambisonics Encoding}
\label{ssec:dps_formulation_AE}

We assume  Ambisonics prior of order $N_p > N_{\text{enc}}$. To account for the heavy-tailed distribution of speech and stabilize the generative modeling process, the prior is defined in a magnitude-compressed \ac{stft} domain~\cite{gerkmann2010empirical,richter2023speech}. We denote this element-wise compression operator as $\compressSTFT{z} = \beta |z|^{\alpha}e^{i \angle(z)}$, where $\alpha$ and $\beta$ are hyperparameters  and $\angle$ extracts the phase. 


To construct the observation vector $\myVec{y}$, 
we employ the linear encoder in \eqref{eq:linear_encoder} of order $N_p$ to match the prior order, yielding 
    $\myVec{y} = \compressSTFT{\tilde{\myVec{E}}\myVec{p}}$,
where $\myVec{p}$ is the $Q$ captured microphone signals.
We then define the forward degradation operator $\mathcal{A}(\cdot)$, which maps the ideal compressed coefficients $\myVec{x}_0$ to the compressed observation space. 
This operator  models linear decompression, physical spatial sampling, linear re-encoding, and STFT compression:
\begin{equation}
    \mathcal{A}(\myVec{x}_0) = \compressSTFT{\tilde{\myVec{E}}\myVec{V}\uncompressSTFT{\myVec{x}_0}}.
\end{equation}

Following \cite{kim2022guided,moliner2023solving}, we normalize the step size in the likelihood score \eqref{eq:likelihood_guidance} by the $\ell_2$-norm of the current measurement gradient to ensure stable gradient dynamics throughout the reverse sampling process. Namely, we set
   $\eta(\tau) = \frac{\eta'}{\sigma({\tau})|\nabla_{\myVec{x}_\tau}\|\myVec{y} - \mathcal{A}(\hat{\myVec{x}}_0)\|_2^2\|_2}$,
where $\eta'$ is a constant scalar throughout the iterative process. 

\vspace{-0.2cm}

\subsection{Inference}
\label{ssec:inference}

To perform Ambisonics encoding, we sample from the posterior distribution by evaluating the \ac{ode}  \eqref{eq:ODE} under the variance-exploding parameterization of~\cite{karras2022elucidating}, where $f(\myVec{x}_\tau,\tau) = 0$, $g(\tau) = \sqrt{2\tau}$, and $\sigma(\tau) = \tau$. Substituting this formulation and incorporating the conditional score function from \eqref{eq:bayes_decomposition}, we arrive at:
\begin{equation*}
    \mathrm{d}\myVec{x}_\tau = -\sigma(\tau) \left[ \nabla_{\myVec{x}_\tau} \log p_{\tau}(\myVec{x}_\tau) + \nabla_{\myVec{x}_\tau} \log p_{\tau}(\myVec{y} \mid \myVec{x}_\tau )\right] \mathrm{d}\sigma(\tau).
\end{equation*}
We numerically integrate this \ac{ode}  in \eqref{eq:score_approx} and \eqref{eq:likelihood_guidance}, while discretizing the time–noise horizon into $M$ steps using a $\rho$-spaced noise schedule~\cite{karras2022elucidating} which set $\tau$ in the range $[\sigma_{\min}, \sigma_{\max}]$ via:
\begin{equation}\label{eq:edm_noise_schedule}
    \tau_{i} =\sigma_{i} = \Big( \sigma_{\max}^{\frac{1}{\rho}} + \frac{i}{M-1} \Big( \sigma_{\min}^{\frac{1}{\rho}} - \sigma_{\max}^{\frac{1}{\rho}} \Big) \Big)^{\rho}.
\end{equation}
The hyperparameter $\rho$ in \eqref{eq:edm_noise_schedule} controls the curvature of the schedule. The complete sampling loop is detailed in Alg.~\ref{alg:diffau}.

\begin{algorithm}
\caption{\ac{adeps}}
\label{alg:diffau} 
\DontPrintSemicolon

\SetKwInOut{Input}{Input}
\Input{Microphone signals $\myVec{p} $;\\
       Steering matrix $\myVec{V}$ ;
       Trained  denoiser $D_\myVec{\theta}(\cdot)$
       }

Set linear encoder $\tilde{\myVec{E}} = \myVec{V}^H \big[ \myVec{V} \myVec{V}^H + \gamma^2 \myVec{I}_Q \big]^{-1}$ \\
Compute observation $\myVec{y} = \compressSTFT{\tilde{\myVec{E}}\myVec{p}}$ \\
Sample $\myVec{x}_{\tau_{\max}} \sim \mathcal{N}(\myVec{y}, \sigma_{\max}^2 \myVec{I}_{N_p})$ \\
\For{$i = 0$ \KwTo $M-1$}{
   Apply denoiser $\tilde{\myVec{x}}_0 \leftarrow D_\myVec{\theta}(\myVec{x}_{\tau_i}, \sigma_{i})$ \\
    Compute  score $\myVec{s}_i \leftarrow (\tilde{\myVec{x}}_0 - \myVec{x}_{\tau_i}) /\sigma_{i}^2 $ \\ 
    Set guidance $\myVec{s}_{\text{LH}} \leftarrow -\eta(\tau_i) \nabla_{\myVec{x}_{\tau_i}} \|\myVec{y} -\mathcal{A}(\tilde{\myVec{x}}_0)\|_2^2$ \\
    Update $\myVec{x}_{\tau_{i+1}} \leftarrow \myVec{x}_{\tau_i} - \sigma_i\big(\sigma_{i+1} - \sigma_i\big) \left[ \myVec{s}_i + \myVec{s}_{\text{LH}} \right]$ 

}

\KwRet{ $\uncompressSTFT{{}\myVec{x}_{0}}$}
\end{algorithm}
\vspace{-0.2cm}

\subsection{Training}
\label{ssec:training}
Algorithm~\ref{alg:diffau} requires a trained diffusion denoiser $D_\myVec{\theta}(\cdot)$. 
We train it on the array-invariant dataset $\mySet{D}$ to learn an ideal order $N_p$ Ambisonics prior. By isolating the network from any specific physical array geometries, spatial aliasing, or microphone noise profiles during training, the learned prior remains entirely array-agnostic. 

We adopt the EDM framework \cite{karras2022elucidating}, 
which preconditions the network to maintain a unit signal variance across all noise levels. The parameters $\myVec{\theta}$ are optimized using a weighted $\ell_2$  objective 
\begin{equation}
\mathcal{L}(\myVec{\theta}) =\mathbb{E}_{\myVec{x}_0, \, \myVec{\epsilon}, \, \tilde{\sigma}} \left[ \lambda(\tilde{\sigma}) \left\| D_\myVec{\theta}(\myVec{x}_0 \!+\! \tilde{\sigma} \myVec{\epsilon}, \sigma_\tau)\! -\! \myVec{x}_0 \right\|^2_2 \right],
\end{equation}
where $\tilde{\sigma}$ is  sampled following \eqref{eq:edm_noise_schedule} and $\lambda(\tilde{\sigma})$ is the loss-weighting profile chosen to counteract the preconditioning scale factors.

\vspace{-0.2cm}
 \subsection{Discussion}
 \label{ssec:discussion} 
\ac{adeps} leverages an Ambisonics prior of order $N_p$ to refine linearly encoded Ambisonics. It frames this process as an inverse problem, incorporating the physical signal model from acquisition to linear encoding. As $\myVec{\theta}$ is trained on ideal Ambisonics signals, it is free from device-specific artifacts (\ref{item:c1}--\ref{item:c3}). These artifacts are instead modeled explicitly during inference, allowing the system to invert them in a zero-shot  array-agnostic  manner (\ref{item:c4}). 

\ac{adeps} exploits the fact that  \eqref{eq:microphone_acquisition_model} exhibits an effective \ac{sh} order $N_{\text{eff}}$, which can be leveraged within the degradation model. Selecting $N_p$ is a design choice: matching $N_p$ closely to $N_{\text{eff}}$ yields superior performance due to more accurate modeling of spatial aliasing at a price of a larger model and memory footprint during training. Nevertheless, we demonstrate in Section~\ref{sec:numerical_study} that even when $N_p < N_{\text{eff}}$, \ac{adeps} consistently outperforms existing array-agnostic baselines.

\begin{table*}[htbp]
\centering
\caption{Performance Evaluation for Order-Matched Ambisonics ($N_{\text{eff}} = N_p = 5$) Across Various Array Configurations.}
\label{tab:results_ideal}
\footnotesize
\setlength{\tabcolsep}{1.2pt} 

\resizebox{\textwidth}{!}{
\begin{tabular}{l | cccc | cccc | cccc | cccc | cccc}
\hline
\multirow{2}{*}{\textbf{Method}} & \multicolumn{4}{c|}{\textbf{SISDR [dB] $\uparrow$}} & \multicolumn{4}{c|}{\textbf{Mean Spec. Error [dB] $\downarrow$}} & \multicolumn{4}{c|}{\textbf{Coherence $\uparrow$}} & \multicolumn{4}{c|}{\textbf{ILD Error [dB] $\downarrow$}} & \multicolumn{4}{c}{\textbf{IC Error $\downarrow$}} \\ 
\cline{2-21}
& 4 Mics & 5 Mics & 6 Mics & Aria & 4 Mics & 5 Mics & 6 Mics & Aria & 4 Mics & 5 Mics & 6 Mics & Aria & 4 Mics & 5 Mics & 6 Mics & Aria & 4 Mics & 5 Mics & 6 Mics & Aria \\ 
\hline
Linear & 6.80 & 7.95 & 9.26 & 3.63 & 8.39 & 7.78 & 7.21 & 14.71 & \textbf{0.81} & 0.82 & 0.83 & \textbf{0.76} & 3.61 & 3.04 & 2.77 & 2.44 & 0.12 & 0.11 & 0.10 & 0.11 \\
Param. & 3.80 & 3.92 & 3.91 & 2.86 & 5.85 & 5.79 & 5.77 & \textbf{6.04} & 0.72 & 0.72 & 0.72 & 0.71 & \textbf{0.98} & 0.91 & 0.92 & 0.93 & 0.09 & 0.08 & 0.08 & 0.09 \\
\ac{adeps} & \textbf{11.66} & \textbf{14.18} & \textbf{16.20} & \textbf{10.00} & \textbf{5.15} & \textbf{4.72} & \textbf{4.35} & 6.67 & 0.80 & \textbf{0.85} & \textbf{0.88} & 0.73 & 1.02 & \textbf{0.68} & \textbf{0.53} & \textbf{0.77} & \textbf{0.07} & \textbf{0.04} & \textbf{0.03} & \textbf{0.06} \\
\hline
\end{tabular}%
}
\end{table*}

\begin{table*}[htbp]
\centering
\caption{Performance Evaluation for Full-Order Ambisonics ($N_{\text{eff}} = 15, N_p = 5$) Across Various Array Configurations.}
\label{tab:results_mismatch}
\footnotesize
\setlength{\tabcolsep}{1.2pt} 

\resizebox{\textwidth}{!}{
\begin{tabular}{l | cccc | cccc | cccc | cccc | cccc}
\hline
\multirow{2}{*}{\textbf{Method}} & \multicolumn{4}{c|}{\textbf{SISDR [dB] $\uparrow$}} & \multicolumn{4}{c|}{\textbf{Mean Spec. Error [dB] $\downarrow$}} & \multicolumn{4}{c|}{\textbf{Coherence $\uparrow$}} & \multicolumn{4}{c|}{\textbf{ILD Error [dB] $\downarrow$}} & \multicolumn{4}{c}{\textbf{IC Error $\downarrow$}} \\ 
\cline{2-21}
& 4 Mics & 5 Mics & 6 Mics & Aria & 4 Mics & 5 Mics & 6 Mics & Aria & 4 Mics & 5 Mics & 6 Mics & Aria & 4 Mics & 5 Mics & 6 Mics & Aria & 4 Mics & 5 Mics & 6 Mics & Aria \\ 
\hline
Linear & 6.71 & 7.82 & 9.09 & 3.31 & 9.56 & 9.32 & 8.37 & 10.84 & \textbf{0.82} & \textbf{0.82} & \textbf{0.83} & \textbf{0.76} & 3.65 & 3.11 & 2.84 & 2.39 & 0.12 & 0.11 & 0.10 & 0.12 \\
Param. & 3.73 & 3.76 & 3.78 & 2.26 & 6.29 & 6.61 & 6.73 & \textbf{7.36} & 0.71 & 0.70 & 0.71 & 0.66 & \textbf{0.95} & 2.54 & 0.88 & \textbf{0.91} & \textbf{0.08} & 0.08 & 0.08 & \textbf{0.08} \\
\ac{adeps} & \textbf{9.85} & \textbf{10.57} & \textbf{12.67} & \textbf{4.94} & \textbf{5.72} & \textbf{5.58} & \textbf{5.03} & 7.43 & 0.74 & 0.76 & 0.80 & 0.64 & 1.21 & \textbf{1.05} & \textbf{0.86} & 1.30 & \textbf{0.08} & \textbf{0.07} & \textbf{0.06} & 0.10 \\
\hline
\end{tabular}%
}

\end{table*}

\begin{table}
  \centering
  \caption{Array specific encoders comparison, 4-mic array in the full-order setting ($N_{\text{eff}}=15$). \textbf{Bold} indicates best, and \underline{underline} indicates second-best. 
  }
  \label{tab:data_driven_comp}
  \footnotesize
   \setlength{\tabcolsep}{2.1pt} 
  \begin{tabular}{l c c c c c}
    \toprule
    & \textbf{SISDR [dB]} $\uparrow$ & \textbf{Sp. Er. [dB]} $\downarrow$ & \textbf{Cohere.} $\uparrow$ & \textbf{ILD Er. [dB]} $\downarrow$ & \textbf{IC Er.} $\downarrow$ \\
    \midrule
    Linear          & \phantom{0}6.63 & 11.02 & 0.80 & 3.24 & 0.13 \\
    U-Net & \phantom{0}6.16 & \textbf{5.83} & \textbf{0.83} & 2.33 & 0.11 \\
    \update{Diff-Enc.}  & \textbf{9.52} & 6.01 & \underline{0.78} & \textbf{0.88} & \textbf{0.06} \\
    ADEPS             & \underline{9.23}  & \underline{5.88} & 0.71 & \underline{1.13} & \underline{0.07} \\
    \bottomrule
  \end{tabular}
\vspace{-0.6cm}
\end{table}

\vspace{-0.2cm}

\section{Numerical Study}
\label{sec:numerical_study}
{\bf Setup and methodology:}
Here, we numerically evaluate \ac{adeps}.\footnote{The source code and hyperparameters used are available at  \url{https://github.com/Amitmils/ADEUPS}}. 
We train an $N_p=5$ order Ambisonics prior. To assess generalizability, we evaluate \ac{adeps} under both an idealized order-matched scenario ($N_{\text{eff}}=5$) and a realistic, full-order sound field ($N_{\text{eff}}=15$) that captures the higher-order spatial components present in real acoustic environments. Simulated target \acp{arir} are generated via  HARP  \cite{saini2025harp} for rooms of size $[6, 10]\times[6, 10]\times[2, 3]\,\text{m}$ and $T_{60} \in [0.1, 0.4]\,\text{s}$. Various arrays are randomly positioned with their center within a  $1 \times 1\,\text{m}^2$ region, with number of speakers ($1\text{--}2$), source distances ($0.8\text{--}1.5\,\text{m}$)  and \acp{doa} are also uniformly sampled per scenario. \acp{arir} are convolved with speech from VCTK~\cite{yamagishi2019cstr} (training) and WSJ0~\cite{garofolo1993csr} (evaluation). The dataset contains $20$k training and $1$k evaluation scenes. 
Our backbone extends NCSN++M \cite{lemercier2023analysing} by modifying input/output layers to $(N_p+1)^2$ channels and adding \acl{adaln}~\cite{peebles2023scalable} time conditioning ($30.8\text{M}$ parameters).
We set compression parameters $(\alpha, \beta) = (0.67, 3)$. Training uses noise schedule $(\sigma_{\max}, \sigma_{\min}, \rho) = (80, 2\times 10^{-3}, 10)$. For inference, warm-init reduces $\sigma_{\max}$ to $20$ over $M=150$ reverse steps.

We evaluate on $13$ test arrays: Project Aria~\cite{engel2023project} and, similar to \cite{heikkinen2025gen}, we generate four random irregular array realizations for each $Q \in \{4, 5, 6\}$ (pairwise spacing: $0.02\text{--}0.18\,\text{m}$). The  signals are projected via \eqref{eq:matrix_signal_acquisition} and corrupted with  Gaussian noise at $50\,\text{dB}$ \ac{snr}, totaling $13$k test signals.

Ambisonics coefficients are evaluated up to $N_{\text{enc}} = 1$. We measure: $(i)$ spectral fidelity and spatial quality using log-magnitude spectrum error and magnitude square coherence \cite{heikkinen2025gen}; $(ii)$ perceptual quality via binaural rendering \cite{Avni2013spatial} (Neumann KU100 HRTF) with \ac{erb}-band \cite{mccormack2022parametric} and mean absolute error of \ac{ild} and \ac{ic} cues \cite{faller2004source}; and $(iii)$ overall reconstruction quality via \ac{si-sdr} \cite{le2019sdr}.

\begin{figure}
    \centering
    \vspace{-0.2cm}
    \includegraphics[width=1\linewidth]{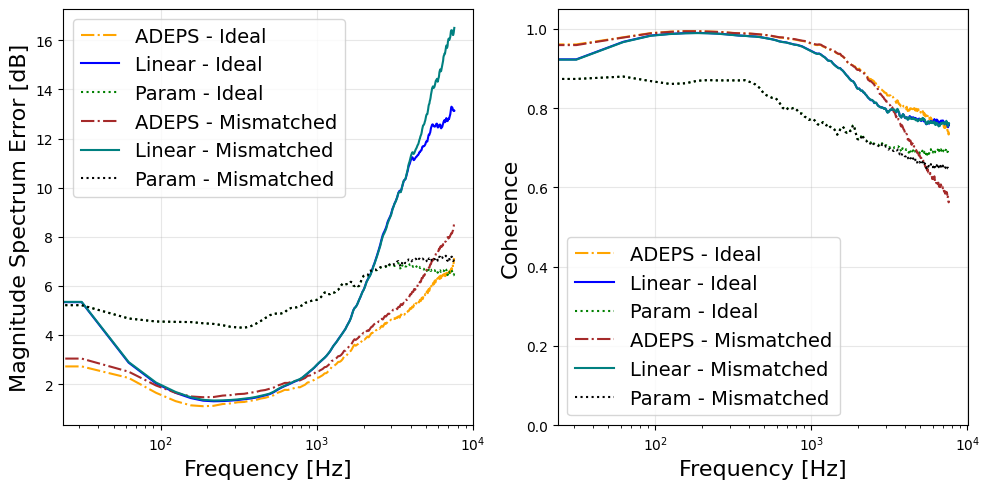}
    \caption{Spectral error and spatial coherence across frequency for idealized and realistic high-order conditions. 
    }
    \label{fig:spectral_comparison}
    \vspace{-0.2cm}
\end{figure}

{\bf Results:} 
We first benchmark \ac{adeps} against arbitrary-array baselines: linear encoding~\cite{moreau20063d} and parametric synthesis~\cite{mccormack2022parametric} (using oracle \acp{doa} and device \acp{atf} for an upper bound). Other data-driven baselines are omitted from here as they cannot handle arbitrary geometries or varying microphone counts. 
As shown in Table~\ref{tab:results_ideal}, \ac{adeps} achieves superior \ac{si-sdr} in the ideal setting and yields the lowest \ac{ild} and \ac{ic} errors.
In the realistic full-order setting (Table~\ref{tab:results_mismatch}), performance degrades as expected as the  model omits higher-order spatial folding. Thus, high-frequency spatial aliasing suppression deteriorates here, causing a drop in coherence relative to linear encoding. Nonetheless, \ac{adeps} maintains a clear lead in \ac{si-sdr} and log-spectral error across all geometries. Further, spectral analysis across all tests (Fig.~\ref{fig:spectral_comparison}) shows that \ac{adeps} consistently reduces low-frequency artifacts (\ref{item:c1}) and suppresses high-frequency spatial aliasing (\ref{item:c3}).

\update{
Finally, Table~\ref{tab:data_driven_comp} compares \ac{adeps} with two array-specific baselines trained on a 4-microphone array
: the U-Net encoder from \cite{heikkinen2024neural} (10.2M parameters) and a diffusion encoder conditioned via channel concatenation (30.8M parameters). \ac{adeps} achieves competitive performance zero-shot without prior exposure to the array geometry. Conversely, the baselines are constrained to a fixed array size and fail completely when evaluated on unseen 4-microphone geometries.}


\vspace{-0.5cm}
\section{Conclusions}
\vspace{-0.2cm}

\label{sec: Conclusions}
\update{We proposed \ac{adeps}, a data-driven, array-agnostic Ambisonic encoder. \ac{adeps} assumes a finite order to construct a spatial prior, which is used to solve the Ambisonics inverse problem via \ac{dps}. Under realistic, high-order sound fields, \ac{adeps} outperforms array-agnostic methods and achieves comparable performance to array-specific baselines, without requiring geometry-dependent training. Currently, the framework is restricted to spatially resolvable orders. A primary objective for future work is extending this method to upscale underdetermined higher-order coefficients.}

 

\begingroup
\footnotesize 
\bibliographystyle{IEEEtran}
\bibliography{IEEEabrv,mybib}

@STRING{IEEE_J_ASSP       = "{IEEE} Trans. Acoust., Speech, Signal Process."}

@STRING{IEEE_J_ASLP       = "{IEEE/ACM} Trans. Audio, Speech, Language Process."}

@STRING{IEEE_M_SP         = "{IEEE} Signal Process. Mag."}

@article{Avni2013spatial,
    author = {Avni, Amir and others},
    title = {Spatial perception of sound fields recorded by spherical microphone arrays with varying spatial resolution},
    journal = {The Journal of the Acoustical Society of America},
    volume = {133},
    number = {5},
    pages = {2711-2721},
    year = {2013}
}

@inproceedings{bastine2022ambisonics,
  title={Ambisonics capture using microphones on head-worn device of arbitrary geometry},
  author={Bastine, Amy and Birnie, Lachlan and Abhayapala, Thushara D and Samarasinghe, Prasanga and Tourbabin, Vladimir},
  booktitle={2022 30th European Signal Processing Conference (EUSIPCO)},
  pages={309--313},
  year={2022},
  organization={IEEE}
}

@article{rafaely2005analysis,
  title={Analysis and design of spherical microphone arrays},
  author={Rafaely, Boaz},
  journal={IEEE Transactions on speech and audio processing},
  volume={13},
  number={1},
  pages={135--143},
  year={2005},
  publisher={IEEE}
}

@book{rafaely2015fundamentals,
  title={Fundamentals of spherical array processing},
  author={Rafaely, Boaz},
  volume={8},
  year={2015},
  publisher={Springer}
}

@incollection{robbins1992empirical,
  title={An empirical Bayes approach to statistics},
  author={Robbins, Herbert E},
  booktitle={Breakthroughs in Statistics: Foundations and basic theory},
  pages={388--394},
  year={1992},
  publisher={Springer}
}

@inproceedings{moreau20063d,
  title={3d sound field recording with higher order ambisonics--objective measurements and validation of a 4th order spherical microphone},
  author={Moreau, S{\'e}bastien and Daniel, J{\'e}r{\^o}me and Bertet, St{\'e}phanie},
  booktitle={120th Convention of the AES},
  pages={20--23},
  year={2006}
}

@inproceedings{qiao2025neural,
  title={Neural Ambisonic Encoding For Multi-Speaker Scenarios Using A Circular Microphone Array},
  author={Qiao, Yue and Kothapally, Vinay and Yu, Meng and Yu, Dong},
  booktitle={IEEE International Conference on Acoustics, Speech and Signal Processing (ICASSP)},
  year={2025}
}

@misc{yamagishi2019cstr,
  title={{CSTR VCTK Corpus}},
  author={Yamagishi, Junichi and Veaux, Christophe and MacDonald, Kirsten},
  year={2019}
}

@article{gayer2025ambisonics,
  title={Ambisonics Encoder for Wearable Array with Improved Binaural Reproduction},
  author={Gayer, Yhonatan and Tourbabin, Vladimir and Ben-Hur, Zamir and Alon, David and Rafaely, Boaz},
  journal={arXiv preprint arXiv:2507.04108},
  year={2025}
}

@inproceedings{politis2017comparing,
  title={Comparing modeled and measurement-based spherical harmonic encoding filters for spherical microphone arrays},
  author={Politis, Archontis and Gamper, Hannes},
  booktitle={2017 IEEE Workshop on Applications of Signal Processing to Audio and Acoustics (WASPAA)},
  pages={224--228},
  year={2017},
  organization={IEEE}
}

@inproceedings{politis2018compass,
  title={{COMPASS}: Coding and multidirectional parameterization of ambisonic sound scenes},
  author={Politis, Archontis and Tervo, Sakari and Pulkki, Ville},
  booktitle={IEEE International Conference on Acoustics, Speech and Signal Processing (ICASSP)},
  pages={6802--6806},
  year={2018}
}

@inproceedings{peebles2023scalable,
  title={Scalable diffusion models with transformers},
  author={Peebles, William and Xie, Saining},
  booktitle={Proceedings of the IEEE/CVF International Conference on Computer Vision (ICCV)},
  pages={4195--4205},
  year={2023}
}

@inproceedings{chung2022diffusion,
  title={Diffusion posterior sampling for general noisy inverse problems},
  author={Chung, Hyungjin and Kim, Jeongsol and Mccann, Michael Thompson and Klasky, Marc Louis and Ye, Jong Chul},
  booktitle={International Conference on Learning Representations (ICLR)},
  year={2022}
}

@article{kim2022guided,
  title={Guided-tts 2: A diffusion model for high-quality adaptive text-to-speech with untranscribed data},
  author={Kim, Sungwon and Kim, Heeseung and Yoon, Sungroh},
  journal={arXiv preprint arXiv:2205.15370},
  year={2022}
}

@article{poletti2005three,
  title={Three-dimensional surround sound systems based on spherical harmonics},
  author={Poletti, Mark A},
  journal={Journal of the Audio Engineering Society},
  volume={53},
  number={11},
  pages={1004--1025},
  year={2005}
}

@article{deppisch2026residual,
  title={Residual Learning for Neural Ambisonics Encoders},
  author={Deppisch, Thomas and Gao, Yang and Mittal, Manan and Stahl, Benjamin and Hold, Christoph and Alon, David and Ben-Hur, Zamir},
  journal={arXiv preprint arXiv:2601.18322},
  year={2026}
}

@inproceedings{moliner2023solving,
  title={Solving audio inverse problems with a diffusion model},
  author={Moliner, Eloi and Lehtinen, Jaakko and V{\"a}lim{\"a}ki, Vesa},
  booktitle={IEEE International Conference on Acoustics, Speech and Signal Processing (ICASSP)},
  year={2023}
}

@article{karras2022elucidating,
  title={Elucidating the design space of diffusion-based generative models},
  author={Karras, Tero and Aittala, Miika and Aila, Timo and Laine, Samuli},
  journal={Advances in neural information processing systems},
  volume={35},
  pages={26565--26577},
  year={2022}
}

@book{zotter2019ambisonics,
  title={Ambisonics: A practical 3D audio theory for recording, studio production, sound reinforcement, and virtual reality},
  author={Zotter, Franz and Frank, Matthias},
  year={2019},
  publisher={Springer Nature}
}

@misc{engel2023project,
  title  = {Project {Aria}: A New Tool for Egocentric Multi-Modal {AI} Research}, 
  author = {Jakob Engel and others},
  year   = {2023},
  eprint = {2308.13561}
}

@article{richter2023speech,
  title={Speech enhancement and dereverberation with diffusion-based generative models},
  author={Richter, Julius and Welker, Simon and Lemercier, Jean-Marie and Lay, Bunlong and Gerkmann, Timo},
  journal=IEEE_J_ASLP,
  volume={31},
  pages={2351--2364},
  year={2023},
  publisher={IEEE}
}

@article{faller2004source,
  author    = {Faller, Christof and Merimaa, Juha},
  journal   = {The Journal of the Acoustical Society of America},
  title     = {Source localization in complex listening situations: Selection of binaural cues based on interaural coherence},
  volume    = {116},
  number    = {5},
  pages     = {3075--3089},
  year      = {2004},
  doi       = {10.1121/1.1791872}
}

@inproceedings{le2019sdr,
  title={{SDR}--half-baked or well done?},
  author={Le Roux, Jonathan and Wisdom, Scott and Erdogan, Hakan and Hershey, John R},
  booktitle={IEEE International Conference on Acoustics, Speech and Signal Processing (ICASSP)},
  pages={626--630},
  year={2019}
}

@inproceedings{saini2025harp,
  title={{HARP}: A large-scale higher-order ambisonic room impulse response dataset},
  author={Saini, Shivam and Peissig, J{\"u}rgen},
  booktitle={IEEE International Conference on Acoustics, Speech, and Signal Processing Workshops (ICASSPW)},
  year={2025}
}

@inproceedings{gerkmann2010empirical,
  title={Empirical distributions of DFT-domain speech coefficients based on estimated speech variances},
  author={Gerkmann, Timo and Martin, Rainer},
  booktitle={Proc. Int. Workshop Acoust. Echo Noise Control},
  year={2010}
}

@inproceedings{lemercier2023analysing,
  title={Analysing diffusion-based generative approaches versus discriminative approaches for speech restoration},
  author={Lemercier, Jean-Marie and Richter, Julius and Welker, Simon and Gerkmann, Timo},
  booktitle={IEEE International Conference on Acoustics, Speech and Signal Processing (ICASSP)},
  year={2023}
}

@misc{garofolo1993csr,
  title        = {{CSR-I (WSJ0)} Complete},
  author       = {Garofolo, John S. and Graff, David and Paul, Douglas B. and Pallett, David S.},
  year         = {1993},
  publisher    = {Linguistic Data Consortium},
  howpublished = {\url{https://catalog.ldc.upenn.edu/LDC93S6A}}
}

@article{song2020score,
  title={Score-based generative modeling through stochastic differential equations},
  author={Song, Yang and Sohl-Dickstein, Jascha and Kingma, Diederik P and Kumar, Abhishek and Ermon, Stefano and Poole, Ben},
  journal={arXiv preprint arXiv:2011.13456},
  year={2020}
}

@inproceedings{heikkinen2024neural,
  title     = {Neural Ambisonics Encoding for Compact Irregular Microphone Arrays},
  author    = {Heikkinen, Mikko and Politis, Archontis and Virtanen, Tuomas},
  booktitle = {IEEE International Conference on Acoustics, Speech and Signal Processing (ICASSP)},
  pages     = {701--705},
  year      = {2024}
}

@inproceedings{heikkinen2025gen,
  title     = {Gen-{A}: Generalizing Ambisonics Neural Encoding to Unseen Microphone Arrays},
  author    = {Heikkinen, Mikko and Politis, Archontis and Drossos, Konstantinos and Virtanen, Tuomas},
  booktitle = {IEEE International Conference on Acoustics, Speech and Signal Processing (ICASSP)}, 
  year      = {2025}
}

@inproceedings{heikkinen2026beyond,
  title     = {Beyond Omnidirectional: Neural Ambisonics Encoding for Arbitrary Microphone Directivity Patterns Using Cross-Attention},
booktitle={ICASSP 2026-2026 IEEE International Conference on Acoustics, Speech and Signal Processing (ICASSP)},
  pages={22587--22591},
  year={2026},
  organization={IEEE}
}

@article{mccormack2022parametric,
  title={Parametric ambisonic encoding of arbitrary microphone arrays},
  author={McCormack, Leo and Politis, Archontis and Gonzalez, Raimundo and Lokki, Tapio and Pulkki, Ville},
  journal=IEEE_J_ASSP,
  volume={30},
  pages={2062--2075},
  year={2022},
  publisher={IEEE}
}

@article{hacihabiboglu2017perceptual,
  title={Perceptual spatial audio recording, simulation, and rendering: An overview of spatial-audio techniques based on psychoacoustics},
  author={Hacihabiboglu, Huseyin and others},
  journal=IEEE_M_SP,
  volume={34},
  number={3},
  pages={36--54},
  year={2017},
  publisher={IEEE}
}
\endgroup

\end{document}